\documentclass[conference]{IEEEtran}
\usepackage{myStyleIEEE}
\usepackage{ushort}
\IEEEoverridecommandlockouts

\def\R{\mathbb{R}}

\usepackage{siunitx}
\AtBeginDocument{\DeclareSIUnit{\MWh}{MWh}}
\usepackage{amsmath}
\usepackage{mathtools}
\usepackage{isomath}

\begin{document}

\title{A Reinforcement Learning Approach for Fast Frequency Control in Low-Inertia Power Systems}

\renewcommand{\theenumi}{\alph{enumi}}

\newcommand{\uros}[1]{\textcolor{magenta}{$\xrightarrow[]{\text{U}}$ #1}}
\newcommand{\vaggelis}[1]{\textcolor{blue}{$\xrightarrow[]{\text{V}}$ #1}}
\newcommand{\pa}[1]{\textcolor{red}{$\xrightarrow[]{\text{P}}$ #1}}

\author{
\IEEEauthorblockN{Ognjen Stanojev\IEEEauthorrefmark{1}, Ognjen Kundacina\IEEEauthorrefmark{2}, Uros Markovic\IEEEauthorrefmark{1}, Evangelos Vrettos\IEEEauthorrefmark{3}, Petros Aristidou\IEEEauthorrefmark{4}, Gabriela Hug\IEEEauthorrefmark{1}}%
\IEEEauthorblockA{\IEEEauthorrefmark{1} EEH - Power Systems Laboratory, ETH Zurich, Switzerland } %
\IEEEauthorblockA{\IEEEauthorrefmark{2} Department of Power, Electronic and Communications Engineering, University of Novi Sad, Serbia
\IEEEauthorblockA{\IEEEauthorrefmark{3} Swissgrid AG, Laufenburg, Switzerland } %
\IEEEauthorblockA{\IEEEauthorrefmark{4} Department of Electrical Engineering, Computer Engineering and Informatics, Cyprus University of Technology, Cyprus} %
\vspace{-0.45cm}
}
}

\maketitle
\IEEEpeerreviewmaketitle

\begin{abstract}
The electric grid is undergoing a major transition from fossil fuel-based power generation to renewable energy sources, typically interfaced to the grid via power electronics. The future power systems are thus expected to face increased control complexity and challenges pertaining to frequency stability due to lower levels of inertia and damping. As a result, the frequency control and development of novel ancillary services is becoming imperative. This paper proposes a data-driven control scheme, based on Reinforcement Learning (RL), for grid-forming Voltage Source Converters (VSCs), with the goal of exploiting their fast response capabilities to provide fast frequency control to the system. A centralized RL-based controller collects generator frequencies and adjusts the VSC power output, in response to a disturbance, to prevent frequency threshold violations. The proposed control scheme is analyzed and its performance evaluated through detailed time-domain simulations of the IEEE 14-bus test system.
\end{abstract}


\vspace{-0.35cm}
\section{Introduction} \label{sec:intro}
The energy sector is in the midst of a fundamental transition reflected in the large-scale integration of Renewable Energy Sources (RES) and the subsequent phase-out of some of the conventional Synchronous Generators (SGs). The shift in generation technology poses new challenges to power system operation due to lower levels of rotational inertia and damping in the system~\cite{Milano2018}. Larger frequency deviations and higher Rate-of-Change-of-Frequency (RoCoF) following a disturbance are thus expected, which can in turn lead to triggering Under Frequency Load Shedding (UFLS) relays and consequent disruptions in the power supply. A new ancillary service - Fast Frequency Control (FFC) - has been proposed in several frequency control areas to meet the upcoming grid challenges~\cite{Meng2020}. Such service can effectively reduce frequency deviations and RoCoF during contingencies by utilizing the available flexibility of fast-responding units. 

Grid-forming Voltage Source Converters (VSCs) with additional battery storage belong to the most suitable unit types for FFC provision due to their fast ramp rates and low response times. Current FFC strategies for \textit{grid-forming} VSCs primarily focus on Virtual Synchronous Machine (VSM) \cite{Zhong2011} and droop-based control techniques \cite{UrosGM} which emulate frequency-power dynamics of synchronous machines. Moreover, the performance of basic VSM and droop controllers can be improved through adaptive parameter (i.e., inertia and damping) tuning \cite{Torres2014,UrosLQR,Saleem2019} or adjustment of controller setpoints \cite{ademolaidowu2019,stanojev2020,Wenlong2020}. In particular, an online optimization algorithm is used in \cite{Torres2014}, a linear quadratic regulator in \cite{UrosLQR} and a neural predictive control scheme in \cite{Saleem2019} to continuously search for optimal parameters during the operation of the VSM. Even though a more efficient limitation of frequency excursions was demonstrated, the inclusion of all relevant operational constraints in the control remains a challenge. The studies in \cite{ademolaidowu2019,stanojev2020,Wenlong2020} employ Model Predictive Control (MPC) methods \cite{MPCbook} to incorporate all control requirements within a single formulation and to produce optimal setpoint adjustments for droop or VSM controllers. Nevertheless, the drawbacks of MPC approaches lie in high computational costs as the optimization needs to be solved online, as well as in considerable observability requirements. 

The recent research efforts in Reinforcement Learning (RL) \cite{Sutton1998} open possibilities for controllers to learn a goal-oriented control law from interactions with a (partially) observable system or its simulation model. Compared to MPC methods, a low online computational effort is required as the optimal control action is obtained by evaluating a function approximator or a look-up table. However, stability and performance certificates are harder to obtain \cite{Ernst2009}. In the domain of power system frequency control, RL methods have previously been considered solely for load frequency control \cite{Ziming2019, Yu2015} and frequency containment \cite{Ebell2018}, with a detailed survey available in \cite{Glavic2019}. In most works \cite{Yu2015,Ebell2018}, the control action is discretized to allow the use of Q-Learning algorithms which in turn might lead to inaccurate control decisions. 

This work develops a centralized RL-based algorithm for FFC that can be integrated as an additional layer to the VSM or droop-based VSC control schemes. It is activated in the events of large active power disturbances, as the last control measure before triggering the UFLS relays. The proposed controller aims to find the optimal adjustment of converters' active power setpoints to keep the frequency nadir and RoCoF within safe operating limits. The optimal control is designed in a data-driven fashion using the Deep Deterministic Policy Gradient (DDPG) algorithm, which is a model-free RL method suitable for continuous control problems. Furthermore, the proposed controller is compared against an equivalent MPC formulation previously developed in \cite{stanojev2020}. We show that significant improvement in online computational efficiency, as well as lower observability requirements, can be achieved at the expense of high offline computational cost. Finally, contrary to \cite{ademolaidowu2019} and \cite{Wenlong2020}, the control performance evaluation is conducted through time-domain simulations using the detailed dynamic model of a low-inertia system from \cite{UrosStability}.

The remainder of the paper is structured as follows. First, theoretical preliminaries on RL and DDPG are introduced in Section~\ref{sec:rl}. Subsequently, Section~\ref{sec:VSC_FFC} motivates the need for FFC and describes the proposed VSC control scheme, whereas Section~\ref{sec:sup_ctrl_design} presents the proposed RL-based algorithm and formulates the equivalent MPC problem. Simulation results and comparison are given in Section~\ref{sec:res}, while Section~\ref{sec:concl} draws the main conclusions of the study.

\section{Theoretical Preliminaries} \label{sec:rl}
\subsection{Reinforcement Learning}
In RL, an agent interacts with a generally stochastic environment at each time step $t$ by making an observation $o_t \in \mathcal{O}$ of the current state of the environment $s_t \in \mathcal{S}$ and applying an action $a_t \in \mathcal{A}$ which governs the environment into a new state $s_{t+1}$. The action is selected based on a policy $\pi: \mathcal{O} \mapsto \mathcal{A}$ such that $a_t = \pi(o_t)$. Additionally, the agent receives a reward signal $r_t \in \mathcal{R}$ defined as a function of the state and the action, i.e., $r: \mathcal{S} \times \mathcal{A} \mapsto \mathcal{R}$. The transition between the environment states is modelled by the transition probability function $p\left(s_{t+1},r_t \,\lvert\, s_t, a_t\right)$, describing the dynamics of the environment. Hence, the RL problem can be expressed as the Markov decision process defined by the 5-tuple $(\mathcal{S}, \mathcal{A}, \mathcal{R}, \mathcal{O}, p)$, where $\mathcal{S}$ is a set of states, $\mathcal{A}$ is a set of actions, $\mathcal{R}$ is a set of immediate rewards, and $\mathcal{O}$ denotes the set of observations.

The objective of the agent is to find the optimal policy that maximizes the expected long-term reward $G_{t} = \sum_{i=0}^{\infty}\gamma^{i}r_{t+i}$ for each time step $t$, with $\gamma \in [0,1]$ representing the discount factor. In finding the optimal policy, many RL algorithms rely on the action-value function ($Q$-function) defined by 
\begin{equation} \label{eq:q_function}
Q^{\pi}(s, a) = \mathop{\mathbb{E}}_{\pi}\,[G_{t} \,\lvert\, s_{t}=s,a_{t}=a],
\end{equation}
where $Q^{\pi}(s, a)$ represents the expected discounted return when taking action $a$ in state $s$ and following policy $\pi$. If the system at hand is not fully observable, the $Q$-function can be defined in terms of observations $o$ instead of states $s$.

Deep Reinforcement Learning (DRL) algorithms attempt to extract the optimal policy from the history of interactions between the agent and the environment using the advances in the deep learning field to approximate the aforementioned action-value and policy functions. In the following we focus on a model-free, off-policy and actor-critic DRL algorithm (i.e., DDPG \cite{Lillicrap2015ContinuousCW}), which demonstrates good results for problems with continuous state and action spaces.

\subsection{Deep Deterministic Policy Gradient} \label{subsec:ddpg}
In the DDPG algorithm \cite{Lillicrap2015ContinuousCW}, two neural networks are used to approximate the $Q$-function \eqref{eq:q_function} and the policy function $\pi$; the so-called critic network $Q(o,a\,\lvert\, \theta^{Q})$ and actor network $\mu(o\,\lvert\, \theta^{\mu})$, respectively, with $\theta^{Q}$ and $\theta^{\mu}$ being the parameters of the respective networks. For a given observation, the actor network outputs an action, while the critic network takes the observation-action pair as the input and outputs the $Q$-value associated with this pair. Moreover, two additional networks - the target actor network $Q'(o,a\,\lvert\,\theta^{Q'})$ and the target critic network $\mu'(o\,\lvert\,\theta^{\mu'})$ - are introduced as time-delayed copies of the actor and the critic networks. They are used for creating the labels for training the original networks, thus making the learning process more stable. In particular, the parameters $\theta^{Q'}$ and $\theta^{\mu'}$ of the target networks slowly track the parameters of the original networks, as follows: 
\begin{subequations}
\label{eq:target_net_updates}
\begin{align}
    \theta^{Q'} &\mapsfrom \tau \theta^{Q} + (1-\tau) \theta^{Q'}, \\
    \theta^{\mu'} &\mapsfrom \tau \theta^{\mu} + (1-\tau) \theta^{\mu'},
\end{align}
\end{subequations}
with the tracking rate defined by the soft update coefficient $\tau \ll 1$.

The training process assumes running a predefined number of episodes, each comprising a series of steps in which the agent interacts with the environment. During the training, the agent explores the environment (i.e., allows less favorable actions  with respect to the current knowledge) by adding the noise sampled from the Ornstein–Uhlenbeck (OU) noise process $\mathcal{N}(\sigma)$ to the actor's output: 
$a_t = \mu(o_t \,\lvert\, \theta^{\mu} )+\mathcal{N}(\sigma)$, where $\sigma$ denotes the OU noise hyperparameter used to quantify the exploration. After each interaction, the tuple $(o_t,a_t,r_t,o_{t+1})$ is stored into the experience replay memory $\mathcal{D}$, from which the minibatches (i.i.d. sets of samples) employed for training of the original networks are selected. 
The critic network is updated by minimizing the loss function, averaged over $N$ samples in the minibatch:
\begin{equation} \label{eq:critic_loss}
L(\theta^{Q}) = \frac{1}{N} \sum_{i=1}^{N}(y_{i} - Q(o_i,a_i\,\lvert\, \theta^{Q}))^2.
\end{equation}
The label for the $i$-th sample in the minibatch is calculated as a sum of the immediate reward received in that sample and the expected $Q$-function value of the next observation $o_i'$, determined by the target actor and critic networks as
\begin{equation} \label{eq:critic_label}
y_{i} = r_{i} + \gamma Q'\left(o_{i}', \mu'(o_{i}' \,\lvert\, \theta^{\mu'}) \,\lvert\, \theta^{Q'}\right).
\end{equation}

The performance of policy $\mu(\cdot\,\lvert\,\theta^\mu)$ can be evaluated for each sample in the minibatch using the policy score function defined by
\begin{equation} \label{eq:policy_score}
    J(\theta^\mu)=\mathop{\mathbb{E}} \big[Q(o,a \,\lvert\, \theta^{Q}) \,\lvert\,  _{o=o_{i},a_{i}=\mu(o_{i})}].
\end{equation}
The policy can be improved by maximizing the policy score function, thus applying the gradient ascent to the actor network. The gradient is approximated by the average value of the policy score function gradients across the minibatch:
\begin{equation} \label{eq:actor_update}
\begin{split}
\nabla_{\theta^{\mu}} J \approx \frac{1}{N} \sum_{i=1}^{N} (\nabla_{a} Q(o_i,a \,\lvert\, \theta^{Q})\,\lvert\,  _{a=\mu(o_{i})} \nabla_{\theta^{\mu}} \mu(o_i \,\lvert\, \theta^{\mu} )).
\end{split}
\end{equation}

\section{Fast Frequency Control Provision by VSCs} \label{sec:VSC_FFC}

\subsection{The Necessity for Fast Frequency Control} \label{subsec:FFC}

Large frequency deviations are presently arrested by the joint effort of Primary Frequency Reserves (PFR) and natural responses of rotating masses of electrical machines and frequency dependent loads. Nevertheless, the current trends in deployment of RES lead to the decommissioning of conventional generation and therefore the reduction in system inertia and oscillation damping. As a result, faster dynamics and larger frequency excursions are expected, contributing to the degradation of system stability margins. The conventional PFR, with typical response times ranging from \SI{10}{\second} to \SI{30}{\second}, are becoming inadequate for containing fast frequency dynamics of low-inertia systems~\cite{Meng2020}. Such concerns necessitate the design of novel frequency services, namely FFC, acting at significantly shorter timescales \cite{Hong2019} (below \SI{1}{s}) to reduce the underlying RoCoF and frequency deviation. Prominent candidates for providing such service are converter-interfaced generators with controllable energy storage due to their fast response times and considerable power capacity. Effective FFC provision by these devices can be enabled by a two-level VSC control scheme comprising inner and outer loops with VSM or droop-based active power control together with a supervisory layer that manages the VSC active power setpoint, as discussed in the following. 

\subsection{FFC Provision by Voltage Source Converters} \label{subsec:vsc_model}

\begin{figure}[!t] 
	\centering
	\scalebox{0.85}{\includegraphics[]{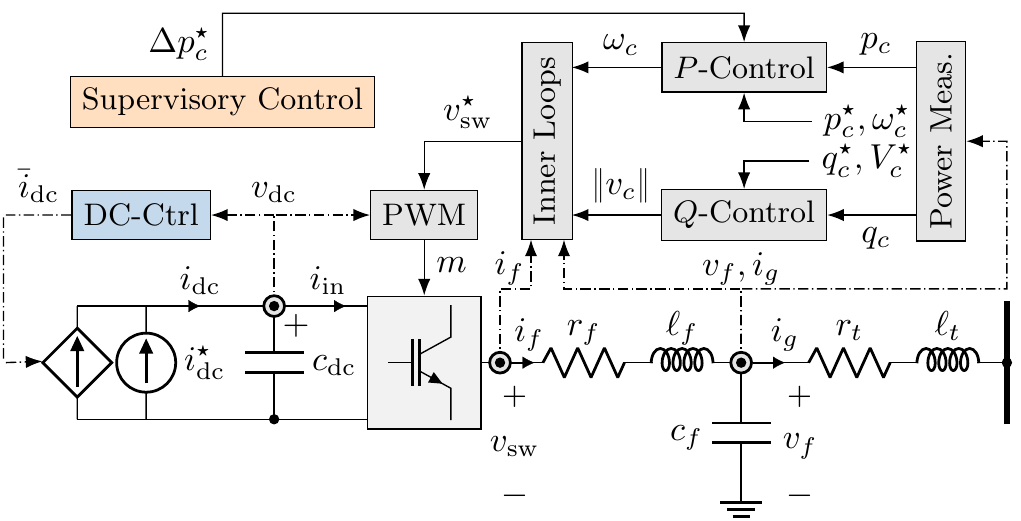}}
	\caption{Proposed control structure, with the added supervisory layer shown in orange. }
	\label{fig:vsc_ctrl_new}
	\vspace{-0.35cm}
\end{figure}

The VSC model considered in this study is composed of a DC-side circuit and lossless switching unit connected to the grid through an $RLC$ filter $(r_f, \ell_f, c_f) \in \mathbb{R}_{>0}^3$ and a transformer $(r_t, \ell_t) \in \mathbb{R}_{>0}^2$, as depicted in Fig.~\ref{fig:vsc_ctrl_new}, with the mathematical model defined in the $dq$-domain and per-unit. The DC-side circuit comprises a DC-link capacitor $c_{\mathrm{dc}} \in \R_{>0}$, a constant current source $i_{\mathrm{dc}}^\star \in \R$ modeling the input of the renewable generation and a controllable DC current source $\bar{i}_{\mathrm{dc}} \in \R$ representing flexibility of the associated energy storage system. A PI controller is employed to maintain the capacitor voltage at a predefined reference $v_\mathrm{dc}^\star\in\R_{>0}$ using the flexibility of the DC-side storage.

The employed VSC control is based on a two-level scheme comprising an inner and an outer control loop. The outer control loop consists of active and reactive power controllers (denoted by $P$- and $Q$-Control in Fig.~\ref{fig:vsc_ctrl_new}) providing the output voltage magnitude $\|v_c\|\in\R$ and frequency $\omega_c\in\R$ references by adjusting the predefined setpoints $(p_c^\star,\omega_{c}^\star,q_c^\star, V_c^\star) \in \R^4$ according to the droop control law and the power measurements $p_c\coloneqq v^\mathsf{T}_f i_g$ and $q_c \coloneqq v^\mathsf{T}_f J i_g$, as follows:
\begin{subequations}
\begin{align}
    &\omega_{c} \coloneqq \omega_{c}^{\star} + R_{c}^p (p_{c}^{\star}+\Delta p_{c}^{\star}-\tilde{p}_c),
   &&\dot{\tilde{p}}_c \coloneqq \omega_f (p_c-\tilde{p}_c), \label{eq:w_c} \\
    &\|v_{c}\| \coloneqq V_{c}^{\star} + R_{c}^q (q_{c}^{\star}-\tilde{q}_c), 
   &&\dot{\tilde{q}}_c \coloneqq \omega_f (q_c-\tilde{q}_c),
\end{align}
\end{subequations}
with $J \in\R^{2\times2}$ denoting the $\ang{90}$ rotation matrix, $v_f \in \R^2$ and $i_g \in \R^2$ representing the filter voltage and the transformer current, respectively; $R_{c}^p \in \mathbb{R}_{>0}$ and $R_{c}^q \in \mathbb{R}_{>0}$ denote the active and reactive power droop gains, $\tilde{p}_c \in \mathbb{R}$ and $\tilde{q}_c \in \mathbb{R}$ represent the low-pass filtered active and reactive power measurements, $\omega_f \in \mathbb{R}$ is the cut-off frequency of the low-pass filter, and $\Delta p_c^{\star} \in \mathbb{R}$ indicates the setpoint change generated by the supervisory layer, discussed in the following section.

The output of the active and reactive power controllers is then passed to the inner control loop comprising a cascade of voltage and current PI controllers, which compute the switching voltage reference $v_\mathrm{sw}^\star\in\R^2$. Note that the modulation voltage reference signal $v_\mathrm{sw}^\star$ is assumed to be perfectly transformed to the AC side, i.e., $v_\mathrm{sw}\coloneqq v_\mathrm{sw}^\star$. Further details on the presented VSC control scheme can be found in \cite{stanojev2020}.

\section{RL-based Supervisory Control Design} \label{sec:sup_ctrl_design}

The goal of the supervisory RL agent is to keep the frequencies of all units in the system after a disturbance within the permissible limits predefined by the TSO by determining appropriate setpoint adjustments $\Delta p_c^\star$. The frequency support in this work is designed as an emergency control scheme, where the RL-based controller remains inactive for acceptable frequency deviations below the threshold. Each FFC providing unit is expected to inject as much power as required to prevent the triggering of UFLS relays in the case of an emergency. 

\subsection{Agent} \label{sec:RL_agent}

In this work, we consider a centralized single agent control architecture, with bidirectional communication links to each VSC $i\in\mathcal{N}_c$ and unidirectional communication lines from each SG $j\in\mathcal{N}_g$ in the system. The communication network is assumed to operate via optical-fibre cables with signal delays below \SI{100}{\milli\second}~\cite{PetrosCyberPhysical2019}. The agent actions are defined as a vector of converter setpoint changes $a = (\Delta p_{c_1}^\star, \dots, \Delta p_{c_{n_c}}^\star) \in \R^{n_c}$ and the agent's observations are frequency and RoCoF measurements at each generator bus $o = (f_1,\dots, f_{n},\dot{f}_1,\dots, \dot{f}_n) \in \R^{2n}$, with $n_c=\lvert\mathcal{N}_c\rvert$ and $n=\lvert\mathcal{N}_c\cup\mathcal{N}_g\rvert$. The RoCoF is measured by sampling the obtained frequency measurements at a rate of \SI{20}{\milli\second} and computing an average over 3 samples. 

\subsection{Training Environment}
The environment for the supervisory RL agent training needs to capture the power system frequency dynamics with sufficient accuracy, while simultaneously being simple enough to alleviate the offline computational burden for the agent training. We use a commonly employed model \cite{GrossIREP,stanojev2020}, where SG dynamics are described by the swing equation and governor control, VSC units by their active power controllers, and network via DC-power flow equations. Hence, a third-order SG model of the form 
\begin{subequations}
\begin{align}
    M_{s_j}\dot{\omega}_{s_j} &= -D_{s_j}\omega_{s_j} + p_{m_j}^\star - p_{s_j},  \label{eq:swing_eq}\\
    T_{g_j}\,\dot{\tilde{p}}_{s_j} &= -\tilde{p}_{s_j} - K_{g_j} \omega_{s_j}, \label{eq:governor}\\
    \dot{\theta}_{s_j}&=\omega_{s_j} \label{eq:angle}
\end{align}
\end{subequations}
is employed, where $x_{s_j} = (\theta_{s_j},\omega_{s_j},\tilde{p}_{s_j})\in\R^3$ is the state vector describing the rotor angle $\theta_{s_j}\in[-\pi,\pi)$, rotor speed $\omega_{s_j}\in\R_{\geq0}$, and dynamics of governor control $\tilde{p}_{s_j}\in\R$ of each synchronous generator $j\in\mathcal{N}_g$; $p_{s_j}\in\R$ indicates changes in the electrical power output, $M_{s_j}\in\R_{>0}$ and $D_{s_j}\in\R_{>0}$ denote generator inertia and damping constants, whereas $T_{g_j}\in\R_{>0}$  and $K_{g_j}\in\R_{>0}$ represent the governor time constant and control gain respectively. The swing equation \eqref{eq:swing_eq} is linearized around a steady state and assumes constant mechanical input $p_{m_j}^\star\in\R_{\geq0}$ over the timescales of interest. A first-order low-pass filter given by \eqref{eq:governor} models the governor dynamics and droop control of the generator \cite{Kundur1994}.
Each VSC-interfaced unit $i\in\mathcal{N}_c$ is modeled with two dynamic states $x_{c_i} = (\theta_{c_i},\tilde{p}_{c_i})\in\R^2$, reflecting the voltage angle $\theta_{c_i}\in[-\pi,\pi)$ and filtered active power $\tilde{p}_{c_i}\in\R$ from \eqref{eq:w_c}. Using droop control, the angle dynamics can be expressed by
\begin{equation} \label{eq:theta_c}
\dot{\theta}_{c_i} = R_{c_i}^p (\Delta p_{c_i}^{\star}-\tilde{p}_{c_i}),
\end{equation}
thus capturing the frequency response of the converter. 

A uniform representation of the network comprising $n_n=\lvert\mathcal{N}_n\rvert$ nodes, $n_b=\lvert\mathcal{N}_b\rvert$ branches, $n_g=\lvert\mathcal{N}_g\rvert$ synchronous and $n_c$ converter-interfaced generators can be established by combining \eqref{eq:theta_c} with \eqref{eq:swing_eq}-\eqref{eq:angle} and the DC power flow balance, resulting in the following linear system:
\begin{subequations}
\label{eq:ss_centralized}
\begin{align} 
    \dot{x} &= A x + B u + E d,\\
    y &= C x + D u + F d,
\end{align}
\end{subequations}
with the system matrices $A\in\R^{(2n_c+3n_g)\times(2n_c+3n_g)}$, $B\in\R^{(2n_c+3n_g)\times n_c}$, $C\in\R^{(n_c+n_g)\times(2n_c+3n_g)}$,
$D\in\R^{(n_c+n_g)\times n_c}$, $E\in\R^{(2n_c+3n_g)\times n_n}$, and $F\in\R^{(n_c+n_g)\times n_n}$, and vectors of variables defined as
{\small
\begin{subequations}
\begin{align}
    x&=\left(x_{c_1},\dots,x_{c_{n_c}},x_{s_1},\dots,x_{s_{n_g}} \right)\in\R^{2n_c+3n_g}, \label{eq:ss_cen_x}\\
    u&=\left(\Delta p_{c_1}^\star,\dots,\Delta p_{c_{n_c}}^\star\right)\in\R^{n_c}, \label{eq:ss_cen_u}\\
    d&=\left(\Delta p_{l_1},\dots,\Delta p_{l_{n_n}}\right)\in\R^{n_n}, \label{eq:ss_cen_d}\\
    y&=\left(f_{c_1},\dots,f_{c_{n_c}},\,f_{s_1},\dots,f_{s_{n_g}}\right)\in \R^{n_c+n_g}.\label{eq:ss_net}
\end{align}
\label{eq:vectors}
\end{subequations}
}%
In \eqref{eq:ss_cen_d}, $\Delta p_{l_k}$ denotes the change in power balance at node $k\in\{1,\dots,n_n\}$, which is treated as a disturbance in this model. Moreover, $f_{s_i}=f_b\omega_{s_i}$ and $f_{c_i}=f_b\omega_{c_i}$  in \eqref{eq:ss_net} represent individual frequencies of SG and VSC units converted into SI, with $f_b=\SI{50}{\hertz}$ being the base frequency. 

Agent actions now coincide with the vector of control inputs (i.e., $a \equiv u$), and the vector of observations is the vector $y$ augmented by the RoCoF measurements. The reward signal is designed to evaluate the effectiveness of agent control actions. For this purpose, a cost coefficient $C_P \in \R_{>0}$ is assigned to the control actions to achieve the targeted objectives under minimum control effort, i.e., $r = - C_P \lVert u \rVert$. The reward signal is further reduced by a constant $C_H \in \R_{>0}$ if any of the generator frequencies $f_i \notin [\ushort{f}_{\mathrm{lim}},\widebar{f}_{\mathrm{lim}}]$ or RoCoF measurements $\dot{f}_i \notin [\ushort{\dot{f}}_{\mathrm{lim}},\widebar{\dot{f}}_{\mathrm{lim}}]$, $\forall i = 1, 
\dots, n$ exceed the permissible range. Hence, by maximizing the reward function, the agent determines a policy that will preserve the relevant frequency metrics within limits while simultaneously minimizing the underlying control effort. In a similar fashion, the device-level converter constraints pertaining to active power and state-of-charge can also be included in the agent design. However, these constraints are omitted from the design for simplicity and brevity of presentation.

\subsection{Algorithm}
During the agent training $M$ episodes are repeated, with each episode consisting of the predefined number of steps $T$ referring to instants when agent-environment interaction takes place. The variety of training scenarios, i.e., episodes, is created by selecting a random step disturbance at a random element of vector $d$ defined in \eqref{eq:ss_cen_d}. At the beginning of each step, an action is selected and executed by adding values to the appropriate elements of vector $u$ defined in \eqref{eq:ss_cen_u}, corresponding to each VSC, and solving the system \eqref{eq:ss_centralized}. Subsequently, state observation variables and the reward signal are created, sent to the agent and stored in the experience replay buffer along with the starting observation and the action. The actor and critic network parameters are updated using the minibatches sampled from the experience replay buffer, as described in Section~\ref{sec:rl}. The step is completed by updating the target networks' parameters, as presented in Algorithm~\ref{training_alg}.

\begin{algorithm}[!b]
\caption{DDPG-based FFC agent training}
\label{training_alg}
\begin{algorithmic}[1]
    \State Initialize critic $Q(o,a\,\lvert\,\theta^{Q})$ and actor $\mu(o\,\lvert\, \theta^{\mu})$ networks
    \State Initialize target networks $Q'(o,a\,\lvert\, \theta^{Q'})$ and $\mu'(o\,\lvert\, \theta^{\mu'})$ with original networks' parameters $\theta^{Q}$ and $\theta^{\mu}$
    \State Initialize the experience replay buffer $\mathcal{D}$
    \For {$\mathrm{episode}=1,2,\ldots,M$}
        \State Generate random disturbance value and select a \NoNumber{random node in the system}
        \State Initialize the environment by simulating generated \NoNumber{disturbance}
        \State Send the initial state variables to the agent
        \For {$t=1,2,\ldots,T$}
            \State Select the action using $a_t = \mu(o_t \,\lvert\, \theta^{\mu} )+\mathcal{N}_t$
            \State Execute action and collect the information about \NoNumber{the immediate reward $r_t$ and the next state $s_{t+1}$}
            \State Store tuple $(o_t, a_t, r_t, o_{t+1})$ in $\mathcal{D}$
            \State Sample the minibatch of tuples from $\mathcal{D}$
            \State Create labels for critic network training using \eqref{eq:critic_label}
            \State Update critic network parameters by minimizing \NoNumber{the loss function given in \eqref{eq:critic_loss}}
            \State Update actor network parameters using \eqref{eq:actor_update}
            \State Update target network parameters using \eqref{eq:target_net_updates}
        \EndFor
	\EndFor
\end{algorithmic}
\end{algorithm}

\subsection{Equivalent MPC Formulation}
For comparison purposes, an equivalent MPC formulation of the RL problem can be established following the modeling procedure in \cite{stanojev2020}. Nonetheless, there are several notable distinctions, as the MPC problem has the possibility of enforcing hard constraints. The objective function aims at minimizing the total control effort over the full horizon $k\in\mathcal{H}$ and over all converter units $i\in\mathcal{N}_c$:
\begin{subequations}
\label{eq:optimization2}
\begin{alignat}{3}
    & \underset{u,\eta_f,\eta_r}{\min} && \,\,\sum_{k\in\mathcal{H}} \,\, C_{P} \lVert u \rVert + C_H \left( \lVert\eta_{f} \rVert_\infty +  \lVert\eta_{r} \rVert_\infty \right)\label{eq:objective2}\\
    & \;\mathrm{s.t.} \quad && \forall k\in\mathcal{H}, \forall i \in \mathcal{N}_c, \forall j\in\mathcal{N}_u, \nonumber \\
    &  && x(k+1) = A_d x(k) + B_d u(k) + E_d d(k) \label{eq:states_pred_cen}, \\
    &  && y(k) = C_d x(k) + D_d u(k) + F_d d(k), \label{eq:otputs_pred_cen} \\
    &  && \dot{f}_j(k) = (f_j(k)-f_j(k-1))/T_s, \label{eq:rocof_discrete_cen}\\
    &  && \ushort{f}_{\mathrm{lim}} - \eta_f(k) \leq f_j(k) \leq \widebar{f}_{\mathrm{lim}} + \eta_f(k), \label{eq:frequencyLimit_cen} \\
    &  && \ushort{\Dot{f}}_{\mathrm{lim}} - \eta_r(k) \leq \Dot{f}_j(k) \leq \widebar{\Dot{f}}_{\mathrm{lim}} + \eta_r(k),\label{eq:RoCoFLimit_cen}\\
    &  && \eta_f(k) \geq 0, \eta_r(k) \geq 0, \label{eq:slackConstraints_cen}
\end{alignat}
\end{subequations}
where $\mathcal{N}_u=\mathcal{N}_g\cup\mathcal{N}_c$ denotes the index set of all generators (including both synchronous and converter-interfaced ones) in the system. The prediction model in \eqref{eq:states_pred_cen}-\eqref{eq:otputs_pred_cen} represents the discrete-time counterpart (denoted by subscript $d$) of the state space given by \eqref{eq:ss_centralized}-\eqref{eq:vectors}, with the vector of node injections $d$ in \eqref{eq:ss_cen_u} being populated by PMU measurements of system disturbances and remaining constant throughout the prediction horizon. The RoCoF is calculated for all generators in \eqref{eq:rocof_discrete_cen}. Frequency and RoCoF constraints are enforced on all generators in \eqref{eq:frequencyLimit_cen}-\eqref{eq:RoCoFLimit_cen}, whereas non-negativity constraints are imposed on slack variables in \eqref{eq:slackConstraints_cen}. It is clear that the MPC formulation has higher observability requirements compared to the RL method, as PMU measurements of power injections at each bus are required for the prediction model in addition to state variable measurements for each generator \eqref{eq:states_pred_cen}.

\section{Results} \label{sec:res}

\begin{figure}[!b]
    \centering
    \vspace{-0.35cm}
    \includegraphics[scale=0.625]{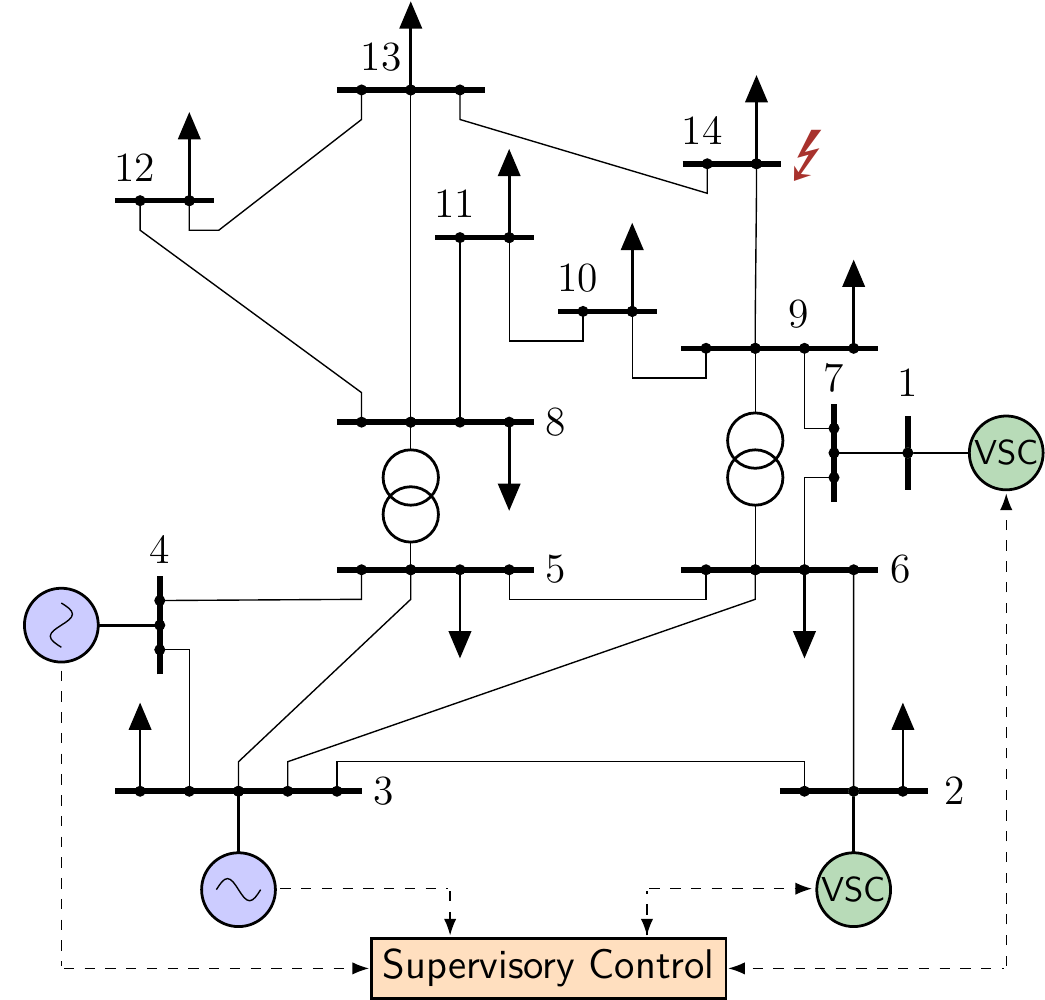}
    \caption{Modified IEEE 14-bus system, with inverter-based generation placed at nodes $1$ and $2$ and indicated communication links between generation units and the supervisory control.}
    \label{fig:14busIEEE_diag}
\end{figure}

All simulations are performed on a modified version of the well-known IEEE 14-bus test system, depicted in Fig.~\ref{fig:14busIEEE_diag}.
The system has been adapted by replacing the SG at node 2 by a VSC unit and including an additional converter-interfaced generator at node 1, with all converter-based units having a power rating of \SI{850}{\mega\watt}. Other system parameters can be found in \cite{Milano2010}. The disturbances considered for control performance evaluation are a loss of generator and a loss of load, both of which can be simulated through step-changes in active power injections at network buses of interest. The triggering of UFLS is assumed to occur in case of frequency deviation beyond $\pm0.5\,\mathrm{Hz}$ or RoCoF magnitudes above $\pm1\,\mathrm{Hz/s}$. Consequently, the limits $f_\mathrm{lim}$ and $\dot{f}_\mathrm{lim}$, used to define the reward function, are set to $50\pm0.5\,\mathrm{Hz}$ and $\pm1\,\mathrm{Hz/s}$, respectively.

\begin{table}[!t]
\renewcommand{\arraystretch}{1.2}
\caption{List of DDPG hyperparameters.}
\label{tab:hyperparameters}
\noindent
\centering
    \begin{minipage}{\linewidth} 
    \renewcommand\footnoterule{\vspace*{-5pt}} 
    \begin{center}
        \begin{tabular}{ l | c }
            \toprule
            \textbf{Hyperparameters} & \textbf{Values} \\ 
            \cline{1-2}
            Experience replay size $\lvert\mathcal{D}\rvert$ & $6\times 10^5$ \\
            \arrayrulecolor{black!30}\hline
            Number of steps $T$ in an episode & $10$ \\
            \arrayrulecolor{black!30}\hline
            Minibatch size $N$ & $256$ \\
            \arrayrulecolor{black!30}\hline
            Discount factor $\gamma$ & $0.99$ \\
            \arrayrulecolor{black!30}\hline
            Actor learning rate & $10^{-5}$ \\
            \arrayrulecolor{black!30}\hline
            Critic learning rate & $10^{-4}$ \\
            \arrayrulecolor{black!30}\hline
            Actor network size (neurons per layer) & $8,128,128,2$ \\
            \arrayrulecolor{black!30}\hline
            Critic network size (neurons per layer) & $10,128,128,128,2$ \\
            \arrayrulecolor{black!30}\hline
            Optimizer & Adam \\
            \arrayrulecolor{black!30}\hline
            Target update factor $\tau$ & $10^{-3}$ \\
            \arrayrulecolor{black!30}\hline
            $\sigma$ parameter for OU noise & $0.02$ \\
            \arrayrulecolor{black}\bottomrule
        \end{tabular}
        \end{center}
    \end{minipage}
\vspace{-0.35cm}
\end{table}

\subsection{Agent Training \& Validation}
We first analyze the agent training process and its performance in the training environment, i.e., the frequency dynamics model from \eqref{eq:ss_centralized}. The discretization step for the state space model is \SI{100}{\nano\second} and the agent-environment interaction takes place every \SI{100}{\milli\second}. The hyperparameters used for DDPG implementation are presented in Table~\ref{tab:hyperparameters}, together with the sizes of actor and critic neural networks. The Rectified Linear Unit (ReLU) activation function is used for all hidden layers of actor and critic networks and subsequently applied to the output of the actor network.

\begin{figure}[!t]
    \centering
    \scalebox{0.91}{\includegraphics{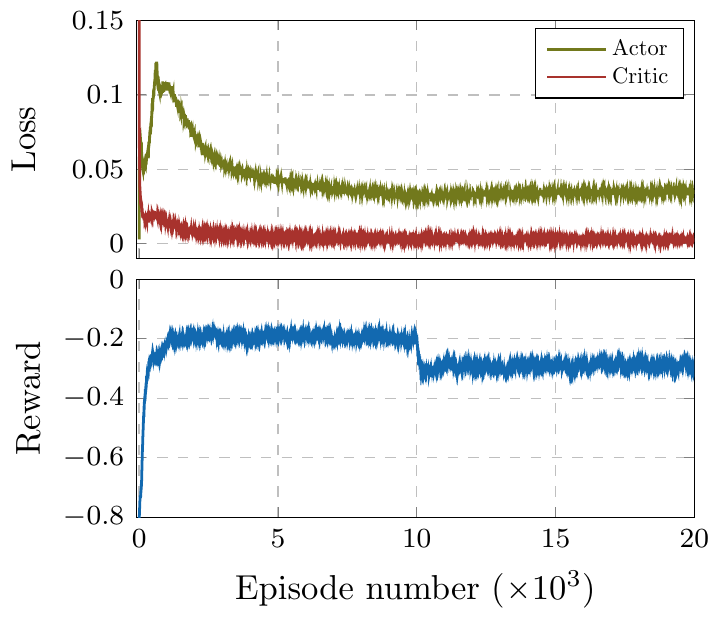}}
    \caption{Average actor and critic loss per episode (top) and total episode reward progress (bottom).}
    \label{fig:tot_episode_reward}
    \vspace{-0.35cm}
\end{figure}

In Fig.~\ref{fig:tot_episode_reward}, the average actor and critic loss per episode are presented. The actor loss is defined as the negative action-value function, since its minimization leads to maximization of the action-value function \eqref{eq:q_function}, and the critic loss is given in \eqref{eq:critic_loss}. As the training progresses, the actor and the critic loss decrease and finally converge to an equilibrium. The initial training shows asymptotic convergence within 8000 episodes. In addition, Fig.~\ref{fig:tot_episode_reward} showcases the progress of the received reward over the trained episodes. To further test the agent adaptiveness to changes in the environment, the inertia constants of generators at buses $3$ and $4$ are reduced by \SI{30}{\percent} after $10^{4}$ episodes. As can be seen from the figure, the agent manages to adapt to the environment change fairly quickly, in approx. 1000 episodes. Understandably, more control effort is required to keep the frequencies within limits for a system with less inertia and hence the total episode reward is lower.

\begin{figure}[!b]
    \centering
    \vspace{-0.35cm}
    \scalebox{0.87}{\includegraphics[width=\linewidth]{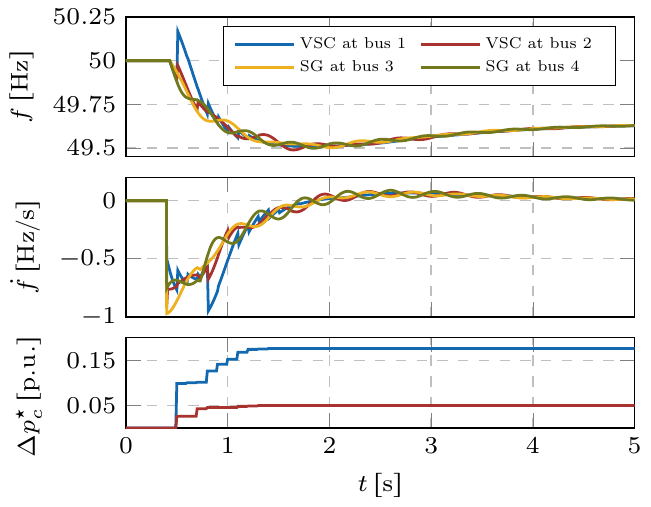}}
    \caption{Frequency and RoCoF responses at each unit bus together with control actions taken by VSCs to keep the frequency and RoCoF within limits.}
    \label{fig:bus14_python_sim}
\end{figure}

After the training is completed, the actor network can be extracted and used to find optimal setpoint changes in real-time operation. To test and verify the agent performance, we apply a disturbance of \SI{800}{\mega\watt} at bus $14$ indicated in Fig.~\ref{fig:14busIEEE_diag}. Generators' frequency and RoCoF responses as well as the applied setpoint changes are depicted in Fig.~\ref{fig:bus14_python_sim}. Provided that the average RoCoF values over \SI{60}{\milli\second} time step are of interest for the RL controller, as discussed in Sec.~\ref{sec:RL_agent}, the RoCoF presented in Fig.~\ref{fig:bus14_python_sim} can be considered a \SI{60}{\milli\second} moving average of the instantaneous RoCoF signal. It can be observed that the contributions of the FFC providing units accurately and timely arrest the frequency decay with minimal control effort. Due to its shorter electrical distance from the fault, the VSC at bus 1 responds at a higher rate compared to other converter-interfaced generator. The instantaneous frequency spikes are a result of the \textit{grid-forming} control strategy.

\subsection{Comparison: RL vs. MPC}
In this section, the system depicted in Fig.~\ref{fig:14busIEEE_diag} is implemented using a high-fidelity dynamic model comprising detailed representations of both synchronous and converter-based generation (see Section~\ref{sec:VSC_FFC}) as well as transmission network dynamics, previously developed in \cite{UrosStability}. The objective is to verify the performance of the RL agent in an environment that closely resembles the real power system and to compare it against the MPC approach from \cite{stanojev2020}.

The prediction horizon of three time steps was chosen in the MPC formulation \eqref{eq:optimization2} to reflect a trade-off between controller performance and computational effort. MPC operates at a time step of \SI{250}{\milli\second} to account for the computational time needed to solve the optimization problem and the communication delays. On the contrary, since evaluation of the actor network is almost instantaneous, the RL agent is chosen to operate every \SI{100}{\milli\second} to account only for the communication delays. A smoother response and a lower control effort of the RL approach are the consequence of the shorter operating time step. 

\begin{figure}[!t]
    \centering
    \includegraphics[width=\linewidth]{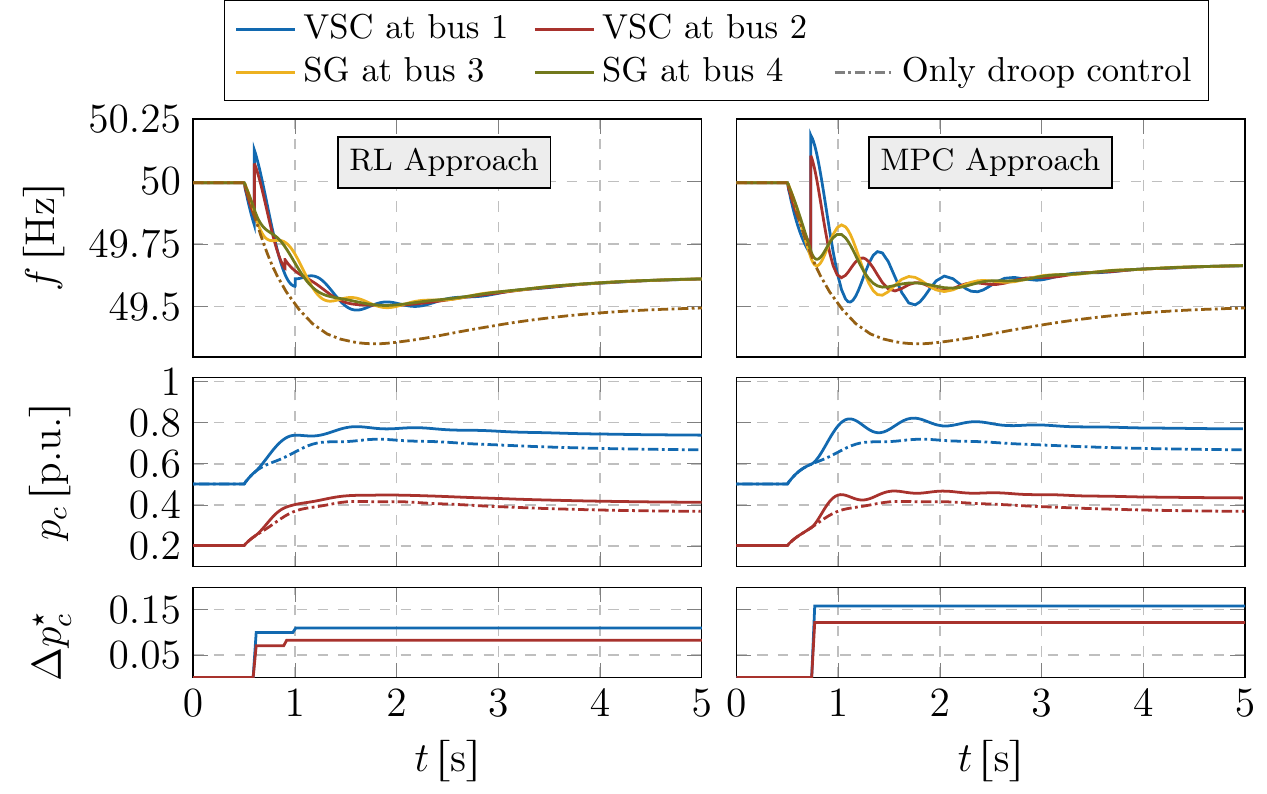}
    \caption{Individual unit frequency, active power and setpoint change responses for the RL (left) and MPC (right) FFC approaches following a disturbance at bus 14. Dashed lines indicate responses of the respective variables for inactive FFC scheme.}
    \label{fig:bus14_comparison}
    \vspace{-0.35cm}
\end{figure}

Fig.~\ref{fig:bus14_comparison} illustrates the MPC and RL agent performances for a disturbance of \SI{1000}{\mega\watt} at bus 14. The dashed frequency response line indicates the center-of-inertia frequency for the case when the FFC scheme is disabled, i.e., only droop control is active. The results suggest that both MPC and RL approaches lead to successful frequency containment, with lower frequency oscillations under the RL agent supervision. Similarly, the dashed lines in the converter power response reflect the VSC activation with droop control only. Hence, the difference between the solid and the dashed lines reflects the contribution of the supervisory layer. A higher control effort is employed under the MPC regulation, resulting in more pronounced oscillations in the VSC power output. 

Note that if the operating time step of the two control schemes was chosen to be the same, the controllers would behave similarly, with RL still being slightly more accurate as it directly maps frequencies to power outputs, without needing the disturbance measurement.

\subsection{Computational Aspects}
The results are obtained on an Intel Core i7-8700 CPU and 26 GB of RAM, with the DAE model implementation done in MATLAB and RL agent training performed in Python using the PyTorch library \cite{PyTorch}. The MPC design has been obtained using YALMIP \cite{Lofberg2004} and CPLEX as the solver. The average computational time needed for determining the optimal setpoints is \SI{150}{\milli\second}. On the other hand, the RL agent obtains the optimal inputs in only \SI{200}{\micro\second}. The difference becomes even more drastic for longer MPC prediction horizons. However, the low online computational burden is shifted to large offline CPU times required for agent training, amounting to \SI{24}{\min} on average for 8000 episodes.

\section{Conclusion} \label{sec:concl}
The paper introduces a novel RL-based FFC scheme for frequency support in low-inertia systems. A supervisory RL agent adjusts power setpoints of \textit{grid-forming} VSCs in response to a disturbance in order to keep the frequency within permissible limits. A DDPG algorithm is used for agent design and training is conducted by simulating a number of random disturbances at arbitrary buses in the system. The training procedure demonstrates fast convergence and capability to quickly adapt to parameter changes in the system. A comparison between the proposed RL-based supervisory controller and a recently developed MPC-based FFC scheme reveals that the RL approach provides smoother frequency control due to shorter activation times and overall lower control effort. Nonetheless, the offline training procedure is required and performance certificates cannot be guaranteed. 

\bibliographystyle{IEEEtran}
\bibliography{bibliography}

\end{document}